\documentclass{article}
\usepackage[english]{babel}
\usepackage[utf8]{inputenc}
\usepackage{johd}
\newcommand{\resultpduration}{$104.8 \pm 8.5$\,ms} 

\title{A bi-atrial statistical shape model for large-scale in silico studies of human atria: model development and application to ECG simulations}

\author{Claudia Nagel$^{a*}$, Steffen Schuler$^{a}$, Olaf Dössel$^{a}$,Axel Loewe$^{a}$ \\
        \small $^{a}$ Institute of Biomedical Engineering, Karlsruhe Institute of Technology, Kaiserstr. 12, 76131 Karlsruhe, Germany \\
        \small $^{*}$Corresponding author: Claudia Nagel, \tt{publications@ibt.kit.edu}
}

\date{} 

\begin{document}

\maketitle

\begin{abstract} 
\noindent Large-scale electrophysiological simulations to obtain electrocardiograms (ECG) carry the  potential to produce extensive datasets for training of machine learning classifiers to, e.g., discriminate between different cardiac pathologies. The adoption of simulations for these purposes is limited due to a lack of ready-to-use models covering atrial anatomical variability.  

We built a bi-atrial statistical shape model (SSM) of the endocardial wall based on 47 segmented human CT and MRI datasets using Gaussian process morphable models. Generalization, specificity, and compactness metrics were evaluated. The SSM was applied to simulate atrial ECGs in 100 random volumetric instances. 

The first eigenmode of our SSM reflects a change of the total volume of both atria, the second the asymmetry between left vs. right atrial volume, the third a change in the prominence of the atrial appendages.
The SSM is capable of generalizing well to unseen geometries and 95\% of the total shape variance is covered by its first 23 eigenvectors. 
The P~waves in the 12-lead ECG of 100 random instances showed a duration of \resultpduration\ in accordance with large cohort studies. 
The novel bi-atrial SSM itself as well as 100 exemplary instances with rule-based augmentation of atrial wall thickness, fiber orientation, inter-atrial bridges and tags for anatomical structures have been made publicly available.

This novel, openly available bi-atrial SSM can in future be employed to generate large sets of realistic atrial geometries as a basis for in silico big data approaches. \end{abstract}

\noindent\\

\section{Introduction}
A wide range of machine learning approaches have already been proposed for classifying cardiovascular pathologies based on the 12-lead electrocardiogram (ECG) \citep{Hannun-2019-ID12428, PerezAlday-2020-ID14773, Strodthoff-2020-ID14032}. Since the ECG is a cost-effective, non-invasive and commonly available tool in clinical practice, it is particularly desirable to identify and diagnose cardiac pathologies only based on the ECG and without the need of further expensive imaging techniques or invasive procedures. However, the training of such classifiers requires a large, balanced, and reliably labeled dataset. Oftentimes, not all of these prerequisites are met when using clinically recorded data. 
Additionally, expert annotations are commonly relied upon to generate the ground truth labels describing the underlying pathologies for clinical datasets coming along with inter- and intra-observer variabilities significantly affecting the reliability of the ground truth labels \citep{Hannun-2019-ID12428}.

These limitations call for simulated synthetic ECG as a source for large, representative and well controlled datasets. These datasets can be used to directly deduce diagnostic criteria visually \citep{Andlauer-2019-ID12169} or to  train machine learning classifiers to discriminate between different cardiac diseases and healthy individuals \citep{Andlauer-2019-ID12169}. The advantage of using simulated over clinical data lies not only in the precisely known ground truth of the underlying pathology that was defined for the simulation, but also in the possibility to generate a virtually infinite amount of signals for each pathology class. 

Nevertheless, atrial, ventricular and thoracic geometrical models are needed for conducting electrophysiological simulations to obtain the 12-lead ECG. 
In this regard, statistical shape models (SSM) allow to compile a wide range of realistic geometries that represent the variability observed in the cohort used to build the SSM. While SSMs of the human ventricles \citep{Bai-2015-ID12225} and torsos \citep{Pishchulin-2017-ID12226}  exist and are publicly available, an open shape model covering both atria covering all relevant anatomical structures for EP simulations (atrial body, appendages, PVs) is still lacking to the best of our knowledge. 

Different statistical atlases of the human whole heart anatomy \citep{ Ecabert-2008-ID14764,Hoogendoorn-2013-ID14760,Lotjonen-2004-ID14761,Ordas-2007-ID14762, Unberath-2015-ID12098, Zhuang-2010-ID14765} have been constructed for segmentation of magnetic resonance (MR) or computed tomography (CT) images by means of active shape modeling approaches. However, those models are usually built based on a small number of sample segmentations or were not made publicly available. 
Furthermore, different SSMs of only the left atrium (LA) have been presented in various studies either for the purpose of simulations \citep{Corrado-2020-ID14628} or for characterizing changes in shape of the LA \citep{Bieging-2018-ID14766,Cates-2014-ID14768,Depa-2010-ID14771,Varela-2017-ID14767} in patients with atrial fibrillation. These LA models are built based on a solid number of instances, but lack the right atrium (RA) and often also the left atrial appendage (LAA). However, these anatomical structures are not only indispensable for the use case of ECG simulations. They are also of particular importance when investigating the mechanisms of typical atrial flutter in the RA or bi-atrial flutter and fibrillation. Additionally, the LAA is highly relevant for studies examining blood clot formation causing stroke \citep{Masci-2019-ID12396}, LAA occlusion as potential therapy \citep{Aguado-2019-ID12516} and the role of the LAA in the onset and maintenance of atrial fibrillation \citep{Nishimura-2019-ID14989}.
Due to the lack of ready-to-use models of the atria, a bi-atrial SSM would cater the need of generating geometrical atrial models representing inter-subject anatomical variability. These could be employed to gain a comprehensive understanding of e.g. the underlying mechanisms of the onset and perpetuation of re-entrant activity during atrial flutter and fibrillation not only in personalized computer models but also in a large patient cohort. 
Thus, including the shape of the RA in the model as well as making the bi-atrial SSM available to the community, enables large-scale simulation of atrial signals. Although the focus of this paper is the application of the SSM for ECG simulations, its field of application is not limited to this particular use case. The bi-atrial model can also be exploited for other in silico approaches like continuum-mechanics and fluid simulations. Furthermore, active shape modeling approaches using the novel bi-atrial SSM could facilitate automated segmentation of the atria from CT or MRI datasets. 

In this work, we built an SSM of both atria from manual segmentations of 47 MR and CT scans. Furthermore, we propose a workflow to generate a volumetric atrial model based on instances of the SSM. We also provide the bi-atrial SSM under the creative commons license CC-BY 4.0 together with 100 exemplary volumetric models derived from it including fiber orientation, inter-atrial bridges and material tags \citep{SSMDatabase}. 

\section{Material and methods}
The geometric representation as well as the variation in shape among a set of individual three dimensional objects can be described by SSMs. 
Point distribution models \citep{cootes95} are the most common subclass of SSMs and require a vectorized point-based representation $\mathbf{s_n}$ of any individual geometry $\mathbf{\Gamma}_n$ comprising $M$ consistently sampled surface points $[\mathbf{x}_n, \mathbf{y}_n, \mathbf{z}_n]^T$:
\begin{equation}
    \mathbf{s}_n = [x_{n,1}, y_{n,1}, z_{n,1}, \dots , x_{n,M}, y_{n,M}, z_{n,M}]^T~.
\end{equation}

Assuming that the spatial variations of the surface points follow a multivariate normal distribution, a compact representation of the mean and covariance matrix describing the shape variations can be obtained by applying a principal component analysis (PCA) to the observations $\mathbf{s}$. In this way, all $N$ individual shapes $\mathbf{\Gamma}$ can be represented by a linear combination of $N-1$ basis functions $\mathbf{v}$: 

\begin{equation}
\label{pca}
    \mathbf{s}_n = \overline{\mathbf{s}} + \sum_{k = 1}^{N-1} r_{n,k} \cdot \sqrt{\sigma^2_k} \cdot \mathbf{v}_k~,
\end{equation}

with $\overline{\mathbf{s}}$ being the mean shape vector as well as $\sigma$ and $\mathbf{v}$ representing the eigenvalues and eigenvectors of the covariance matrix, respectively. $r_{n,k}$ represent the weighting coefficients for the individual eigenvectors.
To obtain this parametric representation of the shape variations from clinical MRI or CT data, a number of preprocessing steps have to be performed: i) segmenting the images, preferably in a (semi-)automatic manner, ii) rigidly aligning the resulting shapes in space, iii) establishing a dense correspondence between the individual shapes to obtain the shape vectors $\mathbf{s}^{C}$ that were then subject to PCA.

\subsection{Dataset}
Three independent multi-center, multi-vendor databases \citep{Karim-2013-ID14757, Karim-2018-ID12272,Tobon-Gomez-2015-ID14763} were used to build the SSM. Their properties are summarized in Table~\ref{tab1e_datasets}. The images originate either from healthy subjects or from patients suffering from atrial fibrillation. Some of the available images were excluded due to low signal to noise ratio or incomplete capture of the inferior right atrial body. Only subjects with 4 pulmonary veins (PVs) were considered because of the limited ability of SSMs to capture only continuous changes of vertex locations. 

\begin{table}[tb]
\caption{\label{tab1e_datasets}Summary of datasets used to generate the statistical shape model.}
\centering
\begin{tabular}{c|p{6cm}|c|l}
\textbf{Dataset} & \textbf{Source} & \textbf{Number of subjects} & \textbf{Voxel resolution}\\
\hline
1& Left atrium segmentation challenge & 30 MRI & 1.25 x 1.25 x 2.7 mm \\
& \citep{Tobon-Gomez-2015-ID14763} &  &  \\
\hline
2& Left atrium fibrosis and scar segmentation challenge & 8 MRI & 1.25 x 1.25 x 2.5 mm \\
 & \citep{Karim-2013-ID14757} &  & 1.4 x 1.4 x 1.4 mm \\
\hline
3& Left atrial wall thickness challenge & 9 CT & 0.8 x 1 x 0.4 mm \\
&  \citep{Karim-2018-ID12272} &  &  \\
\end{tabular}
\end{table}

\subsection{Segmentation}
In order to obtain the individual instances of both atria, a semi-automatic segmentation of the blood pool representing the endocardial surface of the left and the right atrium from MR and CT images was performed using CemrgApp \citep{Razeghi-2020-ID14397}. 2D region growing in several selected slices as well as 3D interpolation were applied to each image stack. To reduce the impact of noise or image artefacts on the segmentation outcome, details were manually corrected.
\begin{figure}[tb]
\centering
\includegraphics[scale=.2]{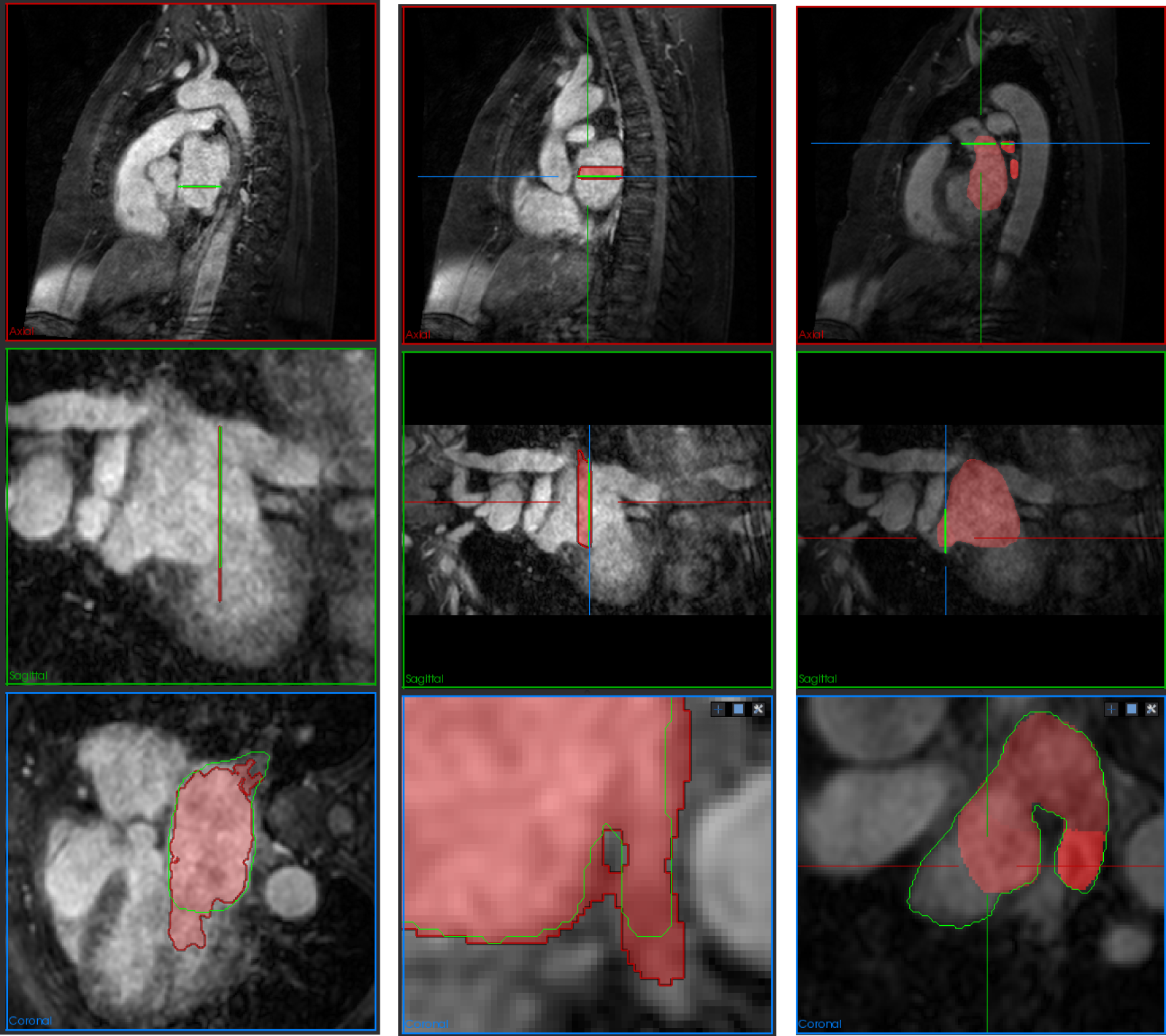}
\caption{\label{segmentation_errors}Segmentation inaccuracy due to different automated segmentation methods. The different rows represent the axial, sagittal and the coronal plane, respectively. The images in the left, middle and right column show the segmentation errors due to region growing, 3D interpolation and a partly excluded LAA in the ground truth data colored in red, respectively. The green contours mark the manual corrections tailored to the correction of inaccurately found automated segmentations.}
\end{figure}
Fig.~\ref{segmentation_errors} shows examples of incorrect segmentation results due to insufficient region growing performance (left column) and 3D interpolation (middle column) that made a manual correction indispensable. Automated segmentation with region growing especially failed in 2D planes where both, the LA and the left ventricle are visible, because the mitral valve shows the same image intensity as the LA and the left ventricle. Therefore, a cutting plane between atrium and ventricle was inserted manually. The drawbacks of 3D interpolation are particularly affecting the areas around the PVs where the interpolated surface tends to close small gaps between the PVs and the atrial body. For 20 images in dataset 2, a segmentation of the LA was provided. However, the LAA was truncated in close proximity to the left atrial body \citep{Tobon-Gomez-2015-ID14763} in these segmentations. Since we aimed at incorporating the variability of the LAA shape in our model, the LA segmentations of dataset 2 were adapted to include the full volume of the LAA as shown in Fig.~\ref{segmentation_errors} (right column). The resulting segmentations were exported as triangular meshes.


\subsection{Rigid Alignment}
After segmenting the individual instances $\mathbf{\Gamma_n}$ of the atria, the different shapes were rigidly aligned in space to avoid a representation of translation and rotation parameters in the eigenmodes of the SSM. This was performed automatically by means of the iterative closest point (ICP) algorithm that provides a solution to the orthogonal Procrustes problem \citep{Chen-1992-ID14942}. In each iteration $i$, candidate correspondences $[\hat{\mathbf{x}}_n, \hat{\mathbf{y}}_n,\hat{\mathbf{z}}_n]_{R,i}^T$ between a target mesh $\mathbf{\Gamma}_n$ and the reference mesh $\mathbf{\Gamma}_{R}$ were found by attributing to each node in $\mathbf{\Gamma}_{n}$ the point with the the smallest Euclidean distance in $\mathbf{\Gamma}_R$. Procrustes analysis was then used to estimate the linear transformation $\mathbf{T}_i$ -- consisting of rotation and translation -- which yields the best match 
of the candidate correspondence points $[\hat{\mathbf{x}}_n, \hat{\mathbf{y}}_n,\hat{\mathbf{z}}_n]_{R,i}^T$
with the reference points $[\mathbf{x}_R, \mathbf{y}_R,\mathbf{z}_R]^T$. 
After applying the estimated transformation $\mathbf{T}_i$ to the points in mesh $\mathbf{\Gamma}_{n}$:

\begin{equation}
\widetilde{\mathbf{\Gamma}}_{{n},{i+1}} = \mathbf{T}_i \cdot \widetilde{\mathbf{\Gamma}}_{{n},{i}}, ~~~ \textrm{with } \widetilde{\mathbf{\Gamma}}_{{n},{0}} = \mathbf{\Gamma}_{n}~,
\end{equation}

new candidate correspondences $[\hat{\mathbf{x}}_n, \hat{\mathbf{y}}_n,\hat{\mathbf{z}}_n]_{R,i+1}^T$ are recursively found between the transformed mesh $\widetilde{\mathbf{\Gamma}}_{{n},{i+1}}$ and $\mathbf{\Gamma}_R$ at each iteration $i$ and used for solving the Procrustes problem. If after several recursive calls of the function, either the maximum number of 150 iterations is exceeded or the convergence criterion is fulfilled, an optimal transformation matrix for the alignment of both shapes $\mathbf{\Gamma}_n$ and $\mathbf{\Gamma}_{R}$ is found resulting in a set of $N$ rigidly aligned shapes ${\mathbf{\Gamma}}^A$ .

\subsection{Establishment of Correspondence}
\begin{figure}
\centering
\includegraphics[scale=.35]{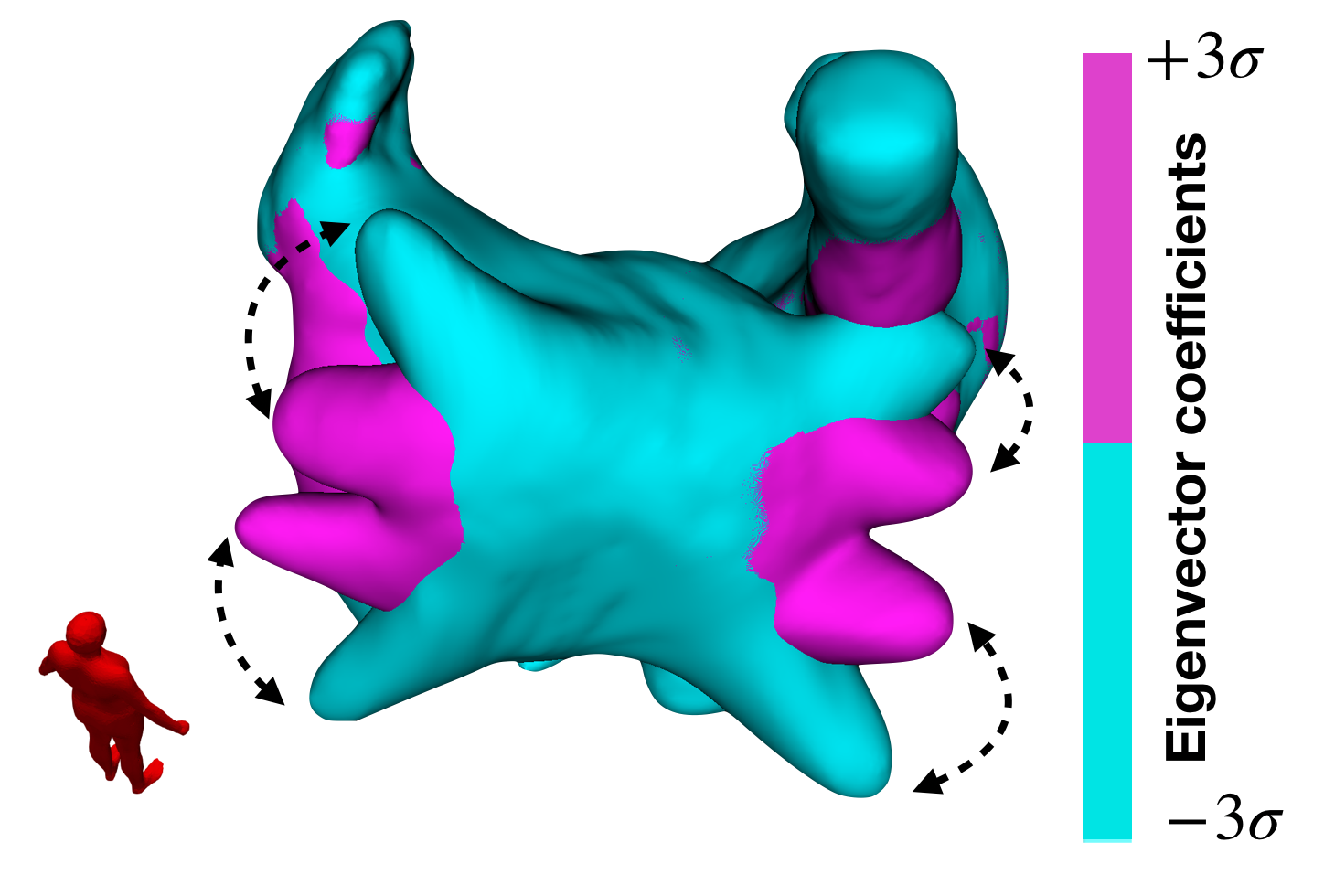}
\caption{\label{pvmodel} Gaussian process morphable model for establishing correspondence at the PVs. The arrows indicate the movement of the respective vertices in the model for a variation of the four leading eigenvectors by $-3\sigma$ and $+3\sigma$.}
\end{figure}

Each aligned individual instance $n$ comprises $M_n$ surface points $[x^A_{n,1}, y^A_{n,1}, z^A_{n,1}, \dots , x^A_{n,M_n}, y^A_{n,M_n}, z^A_{n,M_n}]^T$. In order to describe the variations in shape of the aligned instances ${\mathbf{\Gamma}}^A$ by means of a point distribution model, correspondence between the vertex IDs among all individual shapes have to be established. 
Establishing correspondence requires to retrieve concordant points in all shapes ${\mathbf{\Gamma}}^A$, so that the $N$ aligned shapes are not only represented by the same amount of surface points $M$ but also that each point $[x^A_{n,m}, y^A_{n,m}, z^A_{n,m}]$ with a specific ID $m$ represents the same anatomical landmark in any arbitrary shape $n$. 
For this purpose, we used Gaussian process morphable models (GPMM) \citep{luthi2018} and ScalismoLab \citep{Scalismo:2020} to subject a reference shape ${\mathbf{\Gamma}}^A_R$ to a generic deformation in such a way that the deformed shape $\widetilde{{\mathbf{\Gamma}}}^A_{R,n}$ matches the individual aligned shape ${\mathbf{\Gamma}}^A_n$ in the best possible way. This process then yields a set ${\mathbf{\Gamma}}^{C}$ of aligned shapes that are characterized by homologous, corresponding surface points.
For this process, we defined three independent generic deformations by Gaussian process (GP) models.  Gaussian kernels described the similarity between two distinct points $\mathbf{x}$ and $\mathbf{x'}$:

\begin{equation}
    k(\mathbf{x}, \mathbf{x'}) = s \cdot \exp \left( -\frac{(\mathbf{x}-\mathbf{x'})^2}{l^2} \right)
\end{equation}

were approximated by the leading eigenfunctions of their Karhunen-Loève expansion as described in \cite{luthi2018}. They were further employed to fit the orientation of the left and right pulmonary veins (LPV, RPV), the atrial body, as well as the left and the right atrial appendages (LAA, RAA). The separation into three different models (atrial body, appendages, PVs) served two different purposes. On the one hand, we were able to account for the high anatomical variability of the appendages by allowing smaller inter-dependencies spanning between the points located on the appendages. On the other hand, the generic model varying the points located on the PVs was designed such that only the orientation of the PV ostia but not their length was affected. 
The optimization problem of fitting the GP model $\widetilde{{\mathbf{\Gamma}}}^A_{R,n}$ to the individual aligned shapes ${\mathbf{\Gamma}}^A_n$ was solved with an L-BFGS optimization minimizing the mean squared error between the vertex coordinates of the deformed model $\widetilde{{\mathbf{\Gamma}}}^A_{R,n}$ and the target shape $\mathbf{\Gamma}^A_n$. 
To accurately fit the PVs, a kernel representing the orientation of the four PVs in anterior-posterior direction in its first four eigenvectors was constructed (Fig.~\ref{pvmodel}). Only the orientation of the PVs and not their length were fitted since the latter highly depends on the segmentation approach and does therefore not represent a proper observed anatomical variation considering the heterogeneous input data used in this study. 
A low rank approximation of a GP model with a variance of $s_b = 50, l_b = 40$ at points representing the general atrial body ($b$) and $s_a = 20, l_a = 20$ at the appendages was used to account for the higher anatomical variability in the appendage ($a$) regions. 

\subsection{Shape Model Construction}
\label{ssm_construction}
The vertices at the distal parts of the superior and inferior caval veins, the coronary sinus, the PVs, as well as the mitral valve and the tricuspid valve were discarded from the $N$ aligned shape vectors in correspondence $\mathbf{s}^{C}$ to limit the model creation to atrial components relevant for electrophysiological simulations.
Applying a PCA to these cut shape vectors in correspondence $\mathbf{s}^{C}$ yields their mean shape $\overline{\mathbf{s}}$ and $N-1$ basis functions $\mathbf{v}$ along with their respective variances $\sigma^2$. In this way, an exact reconstruction of any individual shape instance $\mathbf{\Gamma}_n^{C}$ is feasible by determining the coefficients $\mathbf{r}_{n}$ in Eq.~\ref{pca} with a standard least squares estimation. Furthermore, additional arbitrary variation shapes ${\mathbf{\Gamma}}_{var}$ in the span of the $N-1$ basis vectors $\mathbf{v}$ can be derived by varying $\mathbf{r}$. Under the assumption that the values of $\mathbf{r}$ are normally distributed among the observed instances $\mathbf{\Gamma}_n^{C}$, keeping $\mathbf{r}$ in the interval $[-3, +3]^{N-1}$ yields realistic artificially generated shapes within the empirically observed variability.

\subsection{Construction of Volumetric Meshes}
\begin{figure}[tb]
\centering
\includegraphics[width = 0.45\textwidth]{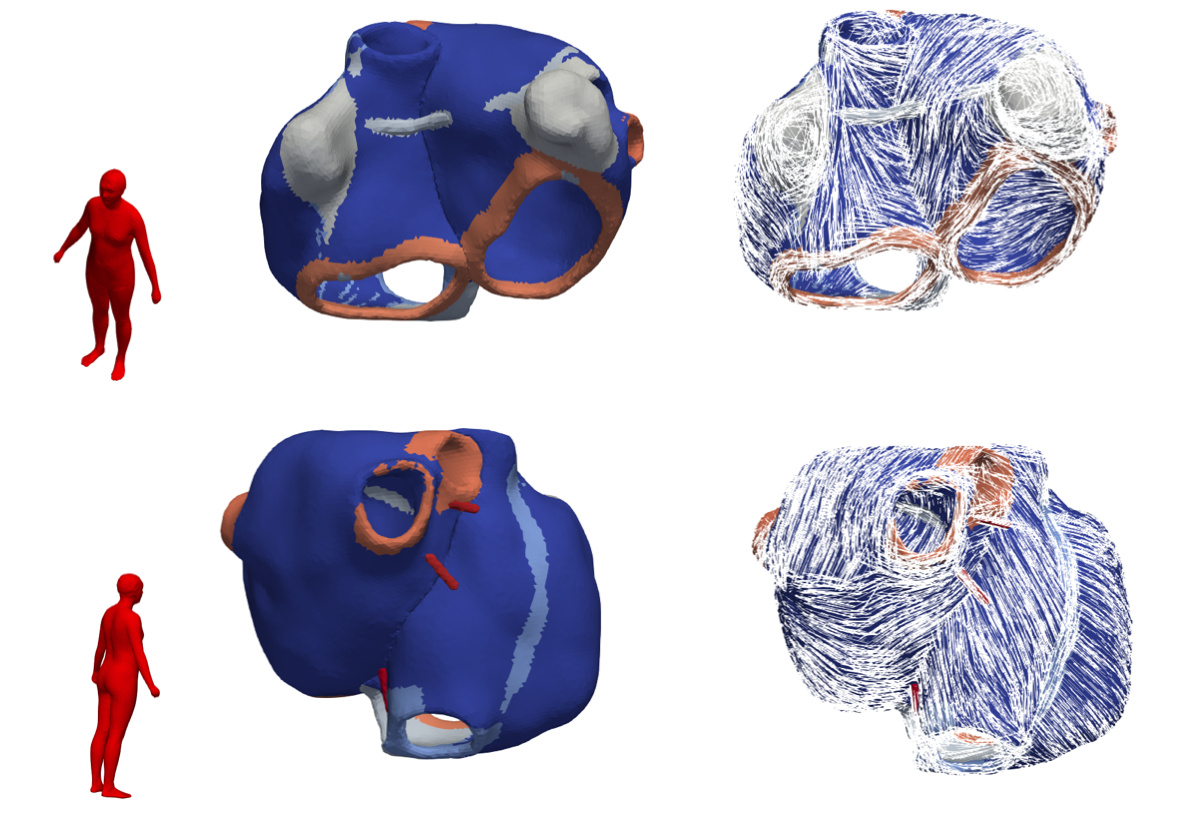}
\caption{\label{volumetricInstances}Example of one randomly generated instance with a homogeneous wall thickness from two different views. The models on the left represent the anatomical labels and the inserted inter-atrial bridges. The fiber orientations are visualized on the models in the right column.}

\end{figure}

The bi-atrial SSM represents the mean shape of the atrial endocardial surface and the variation of all point coordinates in space so that any arbitrary variation mesh ${\mathbf{\Gamma}}_{var}$  can be derived from it. However, a volumetric model of the atria, including inter-atrial bridges, anatomical labels and fiber orientations is required to perform electrophysiological simulations and to obtain realistic body surface P~waves. Since a segmentation of the epicardial surface from conventional MR images is usually not feasible due to an insufficient spatial resolution and a limited signal to noise ratio, the epicardial surface was augmented in a postprocessing step assuming a homogeneous atrial wall thickness.
To approximate the epicardial surface, the endocardial surface of the variation mesh ${\mathbf{\Gamma}}_{var}$ was dilated by 3\,mm at each point along the normal direction calculated as the mean of the adjacent triangle normals. Both surfaces were merged and intersections and holes between epi- and endocardium were corrected and closed automatically using the iso2mesh toolbox \citep{Tran-2020-ID14780}. The closed surface was afterwards remeshed using Instant Meshes \citep{Jakob-2015-ID12648} and transformed into a volumetric tetrahedral mesh with an average edge length of 1\,mm using Gmsh \citep{Geuzaine-2009-ID12650}. The algorithms described by \cite{wachter15,loewe16} were used to automatically augment the models with Bachmann's bundle, a coronary sinus and an upper and middle posterior inter-atrial connection between the LA and RA as well as myocardial fiber orientation and anatomical labels. The augmented anatomical structures are visualized in Fig. \ref{volumetricInstances}.

\subsection{Parameterization of Electrophysiological Simulations}
\label{elphySim}
100 random instances were derived from the bi-atrial SSM by drawing the eigenvector coefficients $\mathbf{r}$ of Eq.~\ref{pca} from a uniform distribution in the range of the minimum and maximum value found in the dataset used to build the SSM. A fast marching \citep{Loewe-2019-ID12386,sethian96} simulation was carried out for solving the Eikonal equation on these 100 geometries to obtain the spread of electrical activation and derive local electrical activation times (LATs) for each node. This sinus rhythm activation was initiated from a sinus node exit site located at the junction of the superior caval vein and the RAA~\citep{loewe16e}. The atrial wall was separated into seven different anatomical regions: regular right and left atrium, inter-atrial connections, valve rings, pectinate muscles, crista terminalis and inferior isthmus. The conduction velocities along the fiber directions and the anisotropy ratios in the different regions were chosen as reported previously \citep{loewe16e} and are given in in Table~\ref{tab1e_FaMaSvalues}. 
\begin{table}[tb]
\centering
\caption{\label{tab1e_FaMaSvalues}Conduction velocity (CV) along the transversal fiber direction in mm/s and anisotropy ratios (AR) in different atrial regions.}
\begin{tabular}{l|l|l}
\textbf{Atrial Region} & \textbf{CV$_T$ (mm/s)} & \textbf{AR} \\
\hline 
Right atrium & 739 & 2.11 \\
Left atrium & 946 & 2.11 \\
Inter-atrial connections & 1093 & 3.36 \\
Valve rings & 445 & 2.11 \\
Pectinate muscles & 578 & 3.78\\
Crista Terminalis & 607    & 3.0 \\
Inferior isthmus & 722 & 1 \\
\end{tabular}
\end{table}
The transmembrane voltage distribution on the atrial surfaces was obtained by shifting a pre-computed \cite{Courtemanche-1998-ID14455} action potential in time according to the calculated LATs as proposed by \citep{kahlmann17a}.
The ECG forward problem to derive the body surface potentials from the transmembrane distribution on the heart was solved by means of the boundary element method \citep{stenroos07}. Considering computational cost, the surface bounding the heart was sampled at a resolution of 3\,mm. Laplacian blurring with optimal blurring parameters as described in \cite{Schuler-2019-ID12702} was applied to the source distribution. 
The mean shape of the human body SSM developed by \cite{Pishchulin-2017-ID12226} served as a reference shape of the torso. The P wave of the 12-lead ECG signal was extracted from the body surface potentials at the standardized electrode positions.

\section{Results}
\subsection{Eigenmodes of the Bi-atrial Shape Model}
\begin{figure}[tb]
\centering
\includegraphics[width = 0.75\textwidth]{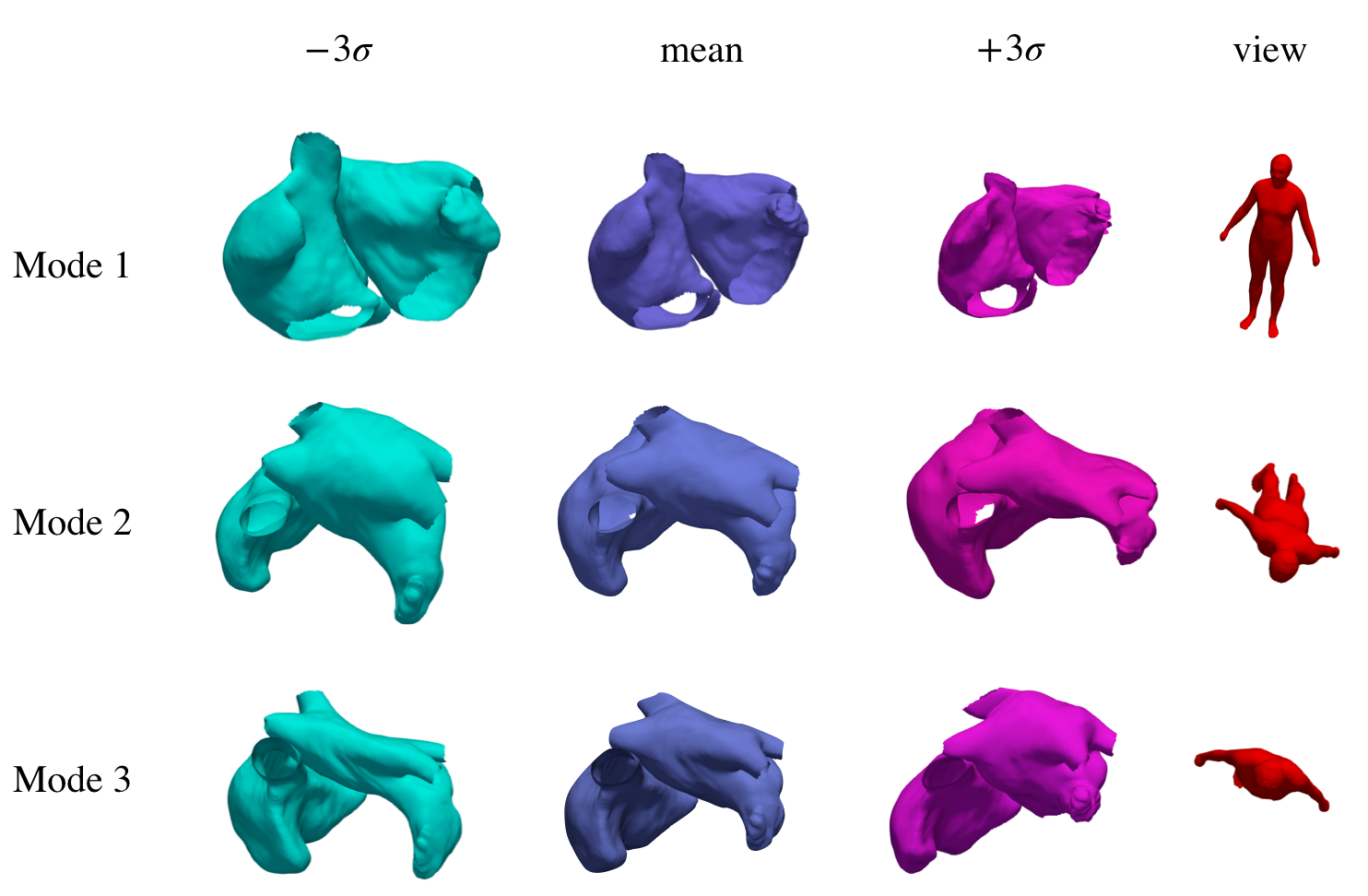}
\caption{\label{eigenmodes}Eigenmodes of the bi-atrial shape model. The three leading eigenmodes are displayed in different rows. Columns 1-3 represent the variation of the eigenvector coefficients. In the fourth column, the anatomical view used to capture the respective eigenmode is depicted.}
\end{figure}

Applying a PCA to the cut shape vectors in correspondence $\mathbf{s}^{C}$ as described in section~\ref{ssm_construction} yields their mean shape $\overline{\mathbf{s}}$ comprising  95.048 triangular cells and 47.528 vertices with an average edge length of 0.862\,mm. Furthermore, the eigenvectors and -values of the bi-atrial model were obtained.
Fig.~\ref{eigenmodes} shows the shape changes caused by varying the coefficients of the first three eigenvectors. The first eigenmode represents a change in the total volume and size of both atria simultaneously (Fig.~\ref{eigenmodes}, first row). The second mode reflects the asymmetry of the LA vs. the RA volume, i.e., the increase of the LA volume and the concurrent decrease of the RA volume. The prominence of the right and left appendage are encoded in the third eigenmode (Fig.~\ref{eigenmodes}, third row). 
The orientation of the PVs to one another are represented -- among other aspects -- in the fifth, sixth, and eighth eigenmodes of the SSM. 

\subsection{Evaluation of the Bi-atrial Shape Model}
The quality of the bi-atrial SSM was evaluated by first assessing the mean vertex to vertex distances between the meshes in correspondence and their respective original locations from the dataset. For the 47 meshes used to build the SSM, this distance was $1.33\pm 0.25$\,mm. Furthermore, three standard evaluation criteria for evaluating the SSM quality proposed by \cite{HuwDavies-2002-ID14807} were considered: generalization, specificity, and compactness. 
The \textit{generalization} metric addressed in section~\ref{generalization} refers to the ability of the SSM to recreate an instance whose shape vector was excluded from the dataset used to create the SSM.
The \textit{specificity} metric (section~\ref{specificity}) assesses the goodness of the model in terms of generating realistic shapes. 
Furthermore, the \textit{compactness} (section~\ref{compactness}) metric of the model increases the more the set of eigenvectors can be reduced while still being able to describe the majority of the total shape variance present in the dataset.

\subsubsection{Generalization}
\label{generalization}
\begin{figure*}[]
\centering
\includegraphics[width = 0.9\textwidth]{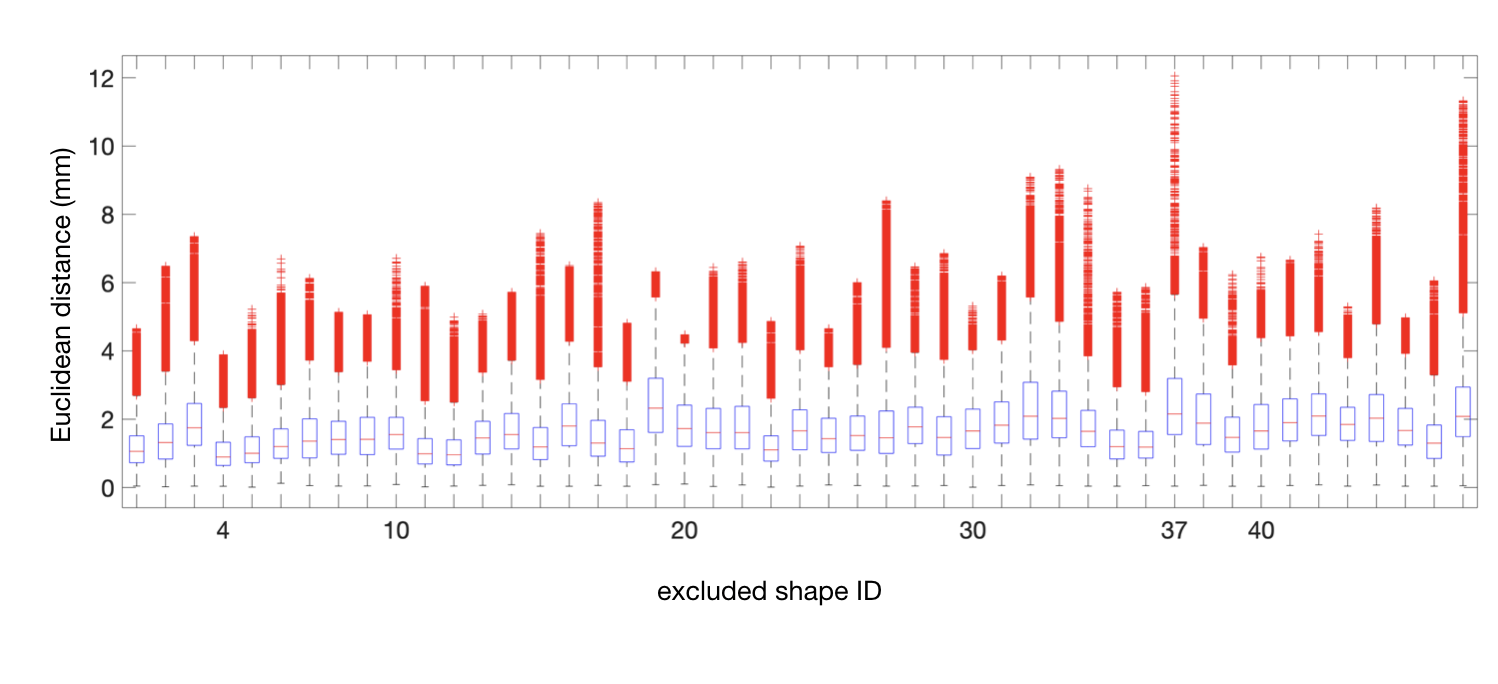}
\caption{\label{generalization_boxplot} Euclidean distance between vertices of the original and reconstructed shapes in mm.}
\end{figure*}

\begin{figure}[tb]
\centering
\includegraphics[width = 0.45\textwidth]{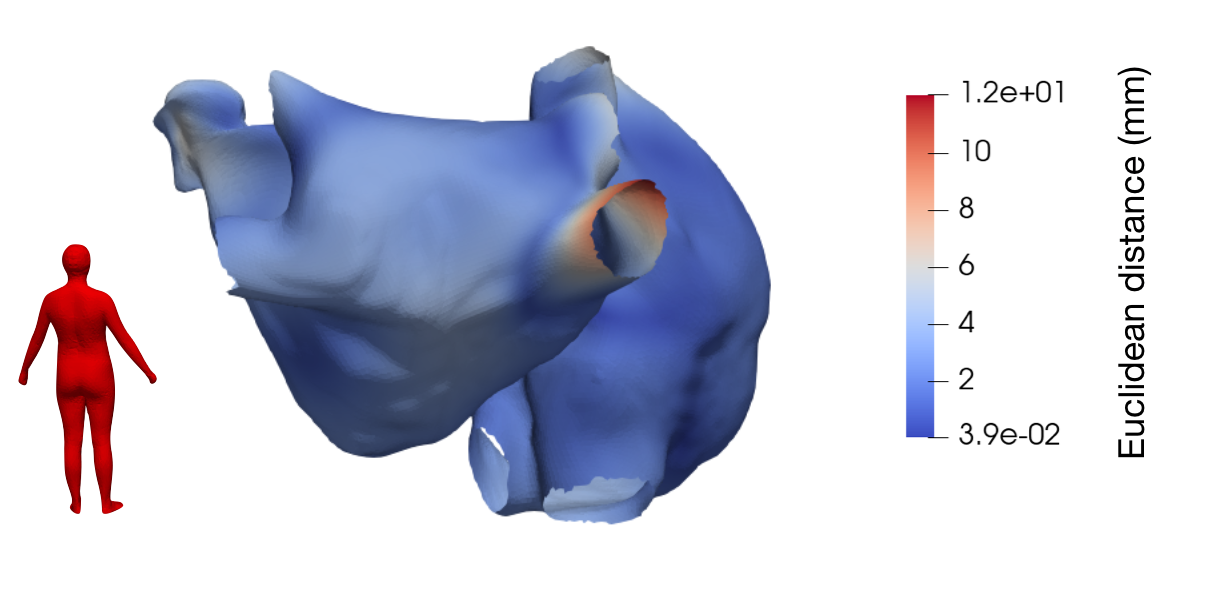}
\caption{\label{generalization_37} Approximated shape of instance 37 with a reduced SSM built without this shape. Vertex color represents the Euclidean distance to the corresponding vertex in the original shape instance 37.}
\end{figure}

To evaluate the generalization quality of the SSM, we used leave-one-out cross-validation and defined $N$ datasets with $N-1$ meshes each by leaving out the final observation. Each of those was used to compute a reduced SSM. The excluded shape was reconstructed with the reduced SSM by determining the eigenvector coefficients using ordinary least squares. The similarity between the original excluded shape and the approximated one was assessed in terms of the Euclidean distance between the corresponding vertices. Fig.~\ref{generalization_boxplot} shows the distribution of this error metric for all instances in the dataset. 
The median of the vertex to vertex distance was below $2.3\,mm$ among all shapes which compares to the order of magnitude of the MRI cross-plane resolution ($0.4\,mm - 2.7\,mm$). Instance 4 holds the lowest Euclidean distances between the vertices of its original and reconstructed shape, whereas instance 37 is characterized by considerably high error values. Especially the 95th percentile bounds and the outlier values comprise large vertex to vertex distances. Fig.~\ref{generalization_37} depicts the approximated atria with the reduced SSM for instance 37. The vertex color represents the Euclidean distance to the corresponding vertex in the original shape instance 37. 
Vertices showing larger deviations were located predominantly on the distal part of the right inferior pulmonary vein (RIPV). The same phenomenon was observed also for instance 47. 

\subsubsection{Specificity}
\label{specificity}
\begin{figure}[tb]
\centering
\includegraphics[width = 0.45\textwidth]{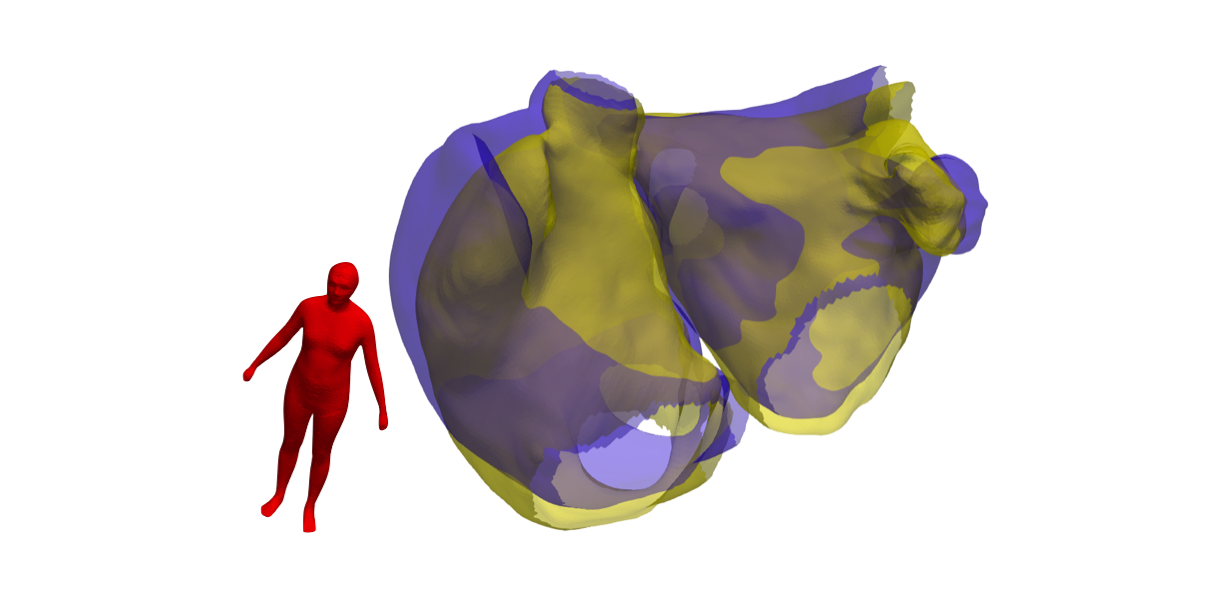}
\caption{\label{spec_example} Example of one randomly generate shape with the SSM (yellow) overlaid with the closest instance in the dataset used to build the SSM (blue).}
\end{figure}

The specificity of the bi-atrial model was evaluated by generating 1000 random shapes according to Eq.~\ref{pca} by uniformly sampling $\mathbf{r}$ in the interval $[-3, 3]$. The similarity between these random shapes and the respective closest shapes in the underlying dataset $\mathbf{\Gamma}^{C}$ used to build the SSM was assessed in terms of the root mean square error (rmse) of all vertex-to-vertex distances between the randomly generated and the original shape. The rmse ranged from $5.29$ to $11.73$\,mm among all 1000 random instances with a mean $\pm$ standard deviation of $7.82\pm 1.04$\,mm. 
Fig.~\ref{spec_example} shows one randomly generated shape with an rmse of $7.58$\,mm in yellow together with its most similar instance from the dataset in blue. This case approximately represents the mean rmse value among all 1000 random shapes. 

\subsubsection{Compactness}
\label{compactness}
\begin{figure}[tb]
\centering
\includegraphics[width = 0.45\textwidth]{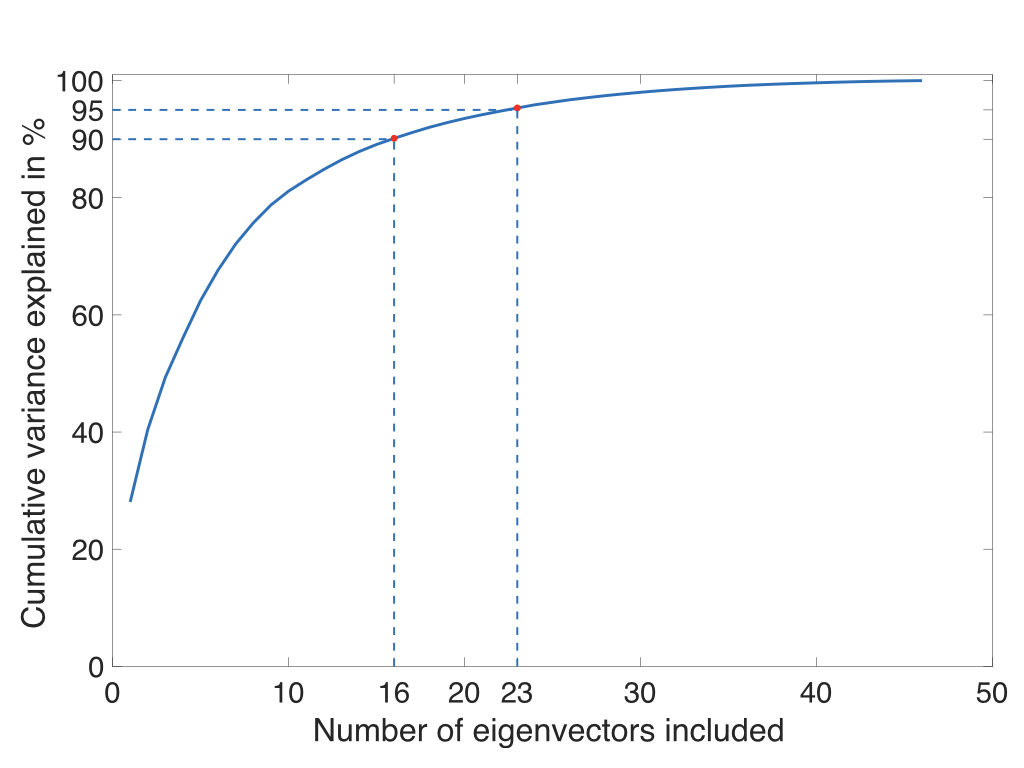}
\caption{\label{convergence_variance} Cumulative variance covered for different numbers of leading eigenvectors included.}
\end{figure}

Fig.~\ref{convergence_variance} depicts the total variance of the dataset explained by the model when including only a limited number of leading eigenvectors. 
90 and 95\% of the total shape variance in the individual segmented shapes can be covered by the SSM when considering only the first 16 and 23 eigenvectors, respectively.
\subsection{P wave Simulations}
\begin{figure*}[tb]
\includegraphics[width = 0.99\textwidth]{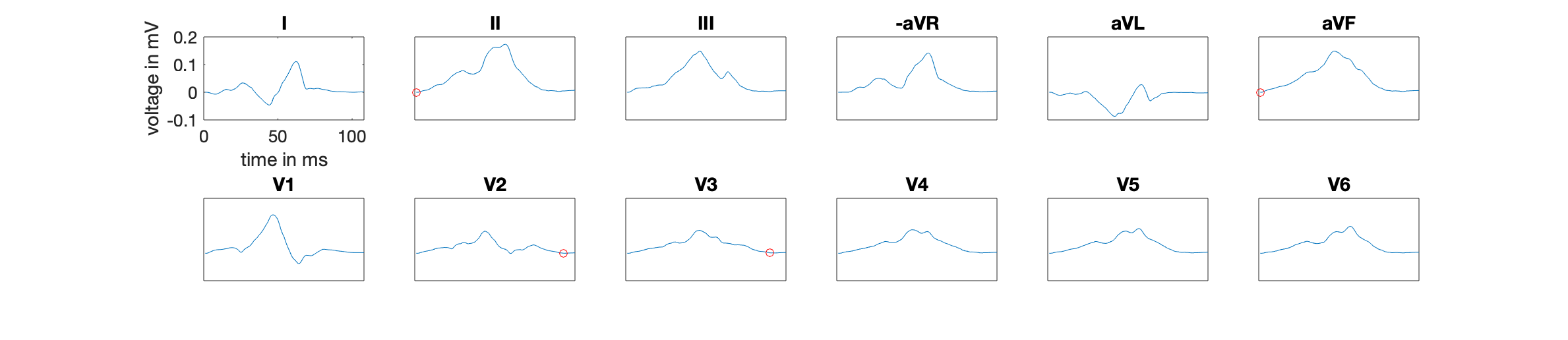}
\centering
\caption{\label{P_meanshape}P~waves of the 12-lead ECG calculated on the mean shape of the proposed bi-atrial shape model. The red markers indicate the earliest onset and the latest offset of the P wave.}
\end{figure*}

Calculating the P wave on the mean shape of the proposed SSM as described in section~\ref{elphySim} yields the signals for the Einthoven, Goldberger and Wilson leads shown in Fig.~\ref{P_meanshape}.
The P wave duration was extracted for each of the 12-lead ECGs simulated on 100 random instances by considering the time difference between the latest detection of the P~wave offset and the earliest P~wave onset across all 12 leads. The values of the P~wave durations  across all 100 instances ranged from 80\,ms to 118\,ms with a mean and standard deviation of \resultpduration. 

\subsection{Openly Available Dataset}
The bi-atrial SSM is provided under the Creative Commons license CC-BY 4.0 together with 100 exemplary volumetric models derived from it including fiber orientation, inter-atrial bridges and anatomical labels \citep{SSMDatabase}.
The SSM is available as an h5 file encoding information about the mean shape's spatial vertex locations and their triangulation. Also the eigenvectors and -values resulting from the applying the PCA are included. 
The 100 geometries were generated by varying the eigenvector coefficients $\mathbf{r}$ in the $[-3, +3] \sigma$ range. These anatomical models are provided in VTK file format including fiber orientation as 3D vectors and material tags as scalar values in the cell data section. 

\section{Discussion}
The main result of this study is a point distribution model incorporating the shape variations of the right and the left atrium as well as their appendages and the PVs. Moreover, we presented a workflow for building a volumetric atrial model from an endocardial surface derived from the SSM. Together with 100 example volumetric geometries generated by varying the coefficients of the principal components uniformly in the $[-3, +3]\sigma$ range including fiber orientation, inter-atrial bridges and anatomical labels, the SSM is openly available \citep{SSMDatabase}. 

Electrophysiological simulations covering atrial excitation spread and propagation of electrical potentials to the body surface were conducted on these 100 example shapes. The resulting P wave durations obtained with the proposed SSM of \resultpduration\ are in agreement with the P wave durations of 100-105\,ms reported for individuals with a low atrial fibrillation risk in an extensive cohort study based on 285,933 ECGs \citep{nielsen15}. On the one hand, this suggests that our model is capable of producing a large variety of variation shapes leading to realistic ECG feature values when compared to clinical recordings. On the other hand, it implies that the additional P wave duration variability observed in individuals with increased atrial fibrillation risk (92-116\,ms range covering 20-80\% percentiles in \cite{nielsen15}) is either due to pathological anatomical variability not represented in the healthy dataset used to build this SSM or due to non-anatomical, functional changes such as conduction velocity slowing due to fibrotic infiltration of the atrial tissue \citep{Caixal-2020-ID14897}.

In our model, 23 eigenvectors are necessary to explain 95\% of the total variance of the dataset. In the LA-only SSM built by \cite{Corrado-2020-ID14628}, the first 15 eigenmodes cover 95\% of the entire shape variance. 
\cite{Cates-2014-ID14768} reported that only 8 eigenvectors account for 95\% of the total variance. However, these two models do neither consider the LAA nor the RA which explains the higher complexity of our model and in turn the need to include more eigenvectors to cover the majority of the shape variance. 
By allowing only a variation of the PVs' orientation in anterior-posterior direction during correspondence retrieval, we prevent changes in the PV lengths and diameters to be reflected in the model's eigenmodes, which was reported as a possible limitation of the model by \cite{Cates-2014-ID14768}. 

Varying the first eigenvector in the LA SSM published by \cite{Varela-2017-ID14767} causes a variation of the total LA volume as it is also the case in our model (Fig.~\ref{eigenmodes}). Also \cite{Corrado-2020-ID14628} and \cite{Cates-2014-ID14768} report a change of the total LA size in the first eigenmode. In the latter, the first mode represents a dilation of the LA mainly in anterior-posterior direction, which is also the case for the third eigenmode of the SSM built by \cite{Corrado-2020-ID14628}, where the first eigenmode rather represents an elongation of the LA in medial-lateral direction. \cite{Cates-2014-ID14768} also constructed a separate SSM of the LAA and found that its first shape parameter corresponds to a change in the LAA length which is as well described by the third eigenmode of our model.

The shape variations encoded in the leading eigenmodes of our novel bi-atrial SSM are consistent with previously published LA-only SSMs as far as the LA is concerned. This demonstrates that our model is able to reflect the same main shape variations even though it is based on a dataset of less than half the size compared to the other models. 
The second eigenmode of our model represents the asymmetry between right and left atrial volume. This further manifests the novel insights into inter-subject atrial geometry variations revealing with our model since this shape change cannot be captured with two separate RA and LA SSMs. 

The generalization results demonstrate that our model is able to accurately predict the shape of a previously unseen atrial geometry. Only the reconstruction of the distal part of the PVs caused vertex-to-vertex distances to the original instances of $>1\,cm$ in two particular cases. The specificity results of $7.82\pm 1.04$\,mm leave room for improvement. However, the low specificity scores do not result from unanatomical characteristics of the generated shapes. They occur rather due to the small dataset of 47 instances available for selecting the closest shape during evaluation. 
Considering the MR slice thickness of predominantly $2.7\,mm$ (Tab.~\ref{tab1e_datasets}), a specificity rmse of $7.82\,mm$ is in the range of less than 2 voxel diameters with segmentation uncertainty adding to it \citep{Karim-2018-ID12272}. The specificity evaluation of our model therefore indicates that randomly generated shapes  produce valid shapes with an accuracy in the range of the error susceptibility during segmentation.

The atria segmented for this study originate from datasets comprising images of not only healthy subjects but also patients with a known history of atrial fibrillation. LA enlargement has been linked with an increased risk for this arrhythmia \citep{Andlauer-2019-ID12169,Broughton-2016-ID14829,Hamatani-2016-ID14830}. To ensure that our model is not based on a biased dataset with predominantly enlarged left atria, the LA volume (excluding LAA and PV ostia) of the $N$ segmented geometries ($82.16\pm19.16\,ml$) was compared to reference values. \cite{Pritchett-2003-ID14831} considered all age and BMI groups in healthy individuals. Translating their 2D measurements to 3D volumes as suggested by \cite{Al-Mohaissen-2013-ID14827} yields a $[-3, +3]\sigma$ range of 10-130\,ml with the largest LA volume in our dataset (122\,ml) being within that range.

To showcase a potential application, we conducted multi-scale electrophysiological simulations on 100 instances of the shape model. This proof of concept was deliberately based on a simple model not considering locally heterogeneous atrial wall thickness \citep{Azzolin-0000-ID14715,Karim-2018-ID12272}, disease-induced remodeling of membrane dynamics \citep{loewe14a} or diffusive aspects during cardiac depolarization. Also only one torso shape and no rotation and translation of the atria within the torso are considered. The pipeline to generate volumetric models and simulation setups from the SSM is prepared for such extensions, though. Future studies focusing on repolarization could replace the simplistic Eikonal coupling employed here with a reaction-Eikonal scheme as suggested by \cite{neic17}.  
%

The main advantage of the novel bi-atrial SSM consists in the automated generation of a large number of atrial geometries. In this way, the cumbersome and time-consuming process of anatomical model generation involving MRI segmentation and defining bundles and fiber orientation can be facilitated and expedited. 

\cite{Potse-2018-ID12698} examined the influence of electrical and structural remodeling on the maintenance of complex reentrant activitiy. With our proposed bi-atrial SSM, also the influence of the general atrial anatomy on the perpetuation and initialization of atrial fibrillation can be quantified. 

\cite{Saha-2016-ID12528} investigated the effect of endo-epicardial activation delay on the P~wave morphology. However, only one atrial geometry was used to deduce models of different atrial wall thicknesses in this study and the authors state the lack of using different geometries as a limitation of their work. The same limitation is listed in the study of \cite{Pezzuto-2018-ID12300} aiming at quantifying the beat-to-beat variability of P~waves in patients with atrial fibrillation. With our SSM, a larger number of different volumetric atrial models is easily acquirable. 

By means of our bi-atrial SSM, scale-large cohort studies using computer models for simulating atrial activity become feasible. \cite{Luongo-0000-ID14752,Luongo-2020-ID14165} found a significant influence of the number of atrial anatomies included on the classification of different types of atrial flutter with a machine learning approach. With the proposed SSM, a large number of geometries can be deduced and used as a basis for in silico big data approaches such as to produce extensive datasets for machine learning applications. The provided instances are ready to be used off the shelf in available computational simulation environments such as openCARP for electrophysiology \citep{Sanchez-2020-ID14887}, openFOAM for fluid dynamics \citep{Jasak-2009-ID14990} or FEniCS for continuum-mechanics \citep{AlnaesBlechta2015a}.



\section{Conclusions}
To the best of our knowledge, we built the first SSM incorporating both atria, their appendages and the orientation of the PVs. The model itself and 100 random volumetric atrial geometries including rule-based fiber orientation and anatomical labelling were made publicly available to the community.  These models are ready to be used off-the-shelf for electrophysiological simulations. 
Established quality criteria indicate that the novel SSM can be reduced to a set of 23 eigenvectors and is capable of generalizing well to   unseen geometries. 
P~waves simulated on 100 random instances derived from the SSM reproduce the P wave distribution observed in clinical ECGs of healthy individuals. As such, the SSM is suitable to generate comprehensive model cohorts covering the relevant anatomical variability as a basis for large-scale in silico simulations including, but not limited to, ECG simulations for machine learning applications.

\section*{Funding Statement}
This work was supported by the EMPIR programme co-financed by the participating states and from the European Union's Horizon 2020 research and innovation programme under grant MedalCare 18HLT07.

\bibliographystyle{apalike}
\bibliography{bib}

\end{document}